\documentclass[11pt,a4paper]{article}
\usepackage[margin=1in]{geometry}
\usepackage{amsmath,amsthm}
\usepackage{newtxtext,newtxmath}
\usepackage{textcomp}
\usepackage{microtype}
\usepackage{booktabs}
\usepackage{array}
\usepackage{graphicx}
\usepackage{placeins}
\usepackage[numbers,sort&compress]{natbib}
\usepackage[hidelinks]{hyperref}
\hypersetup{
  pdftitle={Sliding-Window Quotas in Apache Hadoop YARN: Exact Admission and Flexible Placement},
  pdfauthor={Ruize Zhao}
}
\newtheorem{prop}{Proposition}

\title{Sliding-Window Quotas in Apache Hadoop YARN: Exact Admission and Flexible Placement}
\author{Ruize Zhao\thanks{Corresponding author.} \\ \small Xi\textquotesingle an Jiaotong University, Xi\textquotesingle an, Shaanxi, China \\ \small \texttt{zhaoruize@stu.xjtu.edu.cn}}
\date{}

\begin{document}
\maketitle

\begin{abstract}
\noindent
A reservation in Apache Hadoop YARN can pass a quota of 600 MiB-hours even when a 24-hour window contains 720 MiB-hours of usage. YARN subtracts sampled prefixes to estimate the window total. This can also reject requests that fit, and adding demand can introduce a sample that reverses a rejection. We bound the peak error in both directions and implement an exact check that visits rate changes at either edge of the window. When the two policies see the same history in replays from two public job traces, they disagree on up to 13.3\% of requests. Among requests both policies admit, the 95th-percentile latency of inserting a reservation into the plan is 39--47\% of the sampled baseline. A separate linear scan finds all feasible integer starts for a single constant-demand stage. In a ResourceManager test, the default agent rejects a flexible request. The ResourceManager accepts the stage when submitted at a start found by the scan. In one replay, committing such retries reduces total admissions.
\end{abstract}


\section{Introduction}
Apache Hadoop YARN~\cite{vavilapalli} lets users reserve cluster resources before a job runs. The sharing policy \texttt{CapacityOverTimePolicy}~\cite{yarn1711} limits each user's average allocation over windows of a configured length. At admission, the policy reads the run-length encoded (RLE) timeline of resource rates, constructs two sampled prefix maps, and compares their largest difference with the quota.

Sampling can skip a peak between sample times or carry usage over from before the window starts. The amount carried over can change when the request introduces a new sample. In one regression test, increasing demand lowers the estimated peak from 140 to 84 MiB-seconds while the true peak rises from 80 to 84 MiB-seconds. With a quota of 100 MiB-seconds, the sampled policy admits only the larger request.

The exact check follows changes in rate at both ends of the window. That settles whether a given allocation fits. When the request allows several starts, the agent's choice can exceed quota even though the stage fits elsewhere before its deadline. For a single stage, the same timeline events determine every feasible start.

We derive tight bounds for the unrounded sampled check and use Hadoop 3.5.0 to test an exact replacement at the plan entry point and through the ResourceManager client interface. The Java policy handles multiple resources, replacements and recurring allocations. For plans that already exceed quota, we test a separate rule that allows updates without increasing any window's excess. We also give a linear scan for the feasible starts of one stage. The replays measure how often the validators disagree and how admitting a request at a later start affects the jobs that follow.

\section{Background and Model}
\subsection{Reservation admission}
A reservation agent turns a client's time range and resource requests into an allocation~\cite{rayon,planning,hadoopres}, which the plan checks against its sharing policy before committing it. When the deadline equals the arrival plus the duration of a single stage, the start is fixed. A wider range gives the agent room to move the stage. The CapacityScheduler sets an instantaneous limit on the resource rate and an average limit over a window, with \texttt{CapacityOverTimePolicy}~\cite{capacitydocs} as its default policy.

\subsection{Exact and sampled window totals}
Let $W>0$ be the window length, $Q$ the quota in resource-time units, and consider one resource dimension. Usage $u(t)\in[0,M]$ is piecewise constant, changes at integer timestamps, and is zero before the origin. A reservation of $d$ units on $[a,b)$ adds $d$ to $u$ there. With $P(t)=\int_0^t u(s)\,ds$, the window total is
\[
F(t)=\int_{t-W}^{t}u(s)\,ds=P(t)-P(t-W).
\]
In a compliant plan, inserting $[a,b)$ requires checking $F(t)\le Q$ for $t\in[a,b+W]$, the range of right endpoints for windows whose totals can change. An update removes the old allocation and adds the new one.

The plan provides an RLE map of rates. At validation the sampled policy builds two maps of cumulative usage: it samples the change points, inserts intermediate points $W$ apart inside long segments, and places each prefix value at its sample time and $W$ later, both maps holding the last value until the next key. Subtraction yields, before rounding,
\[
H(t)=P(p(t))-P(p(t-W)),
\]
where $p(x)$ is the latest sample at or before $x$. Before the first sample we take $p(x)=x$ and the prefix is zero. YARN divides each prefix by $W$ and rounds it before subtraction.

\subsection{A missed 24-hour window}
\label{sec:missed-peak}
\begin{figure}[tb]
\centering
\includegraphics[width=\linewidth]{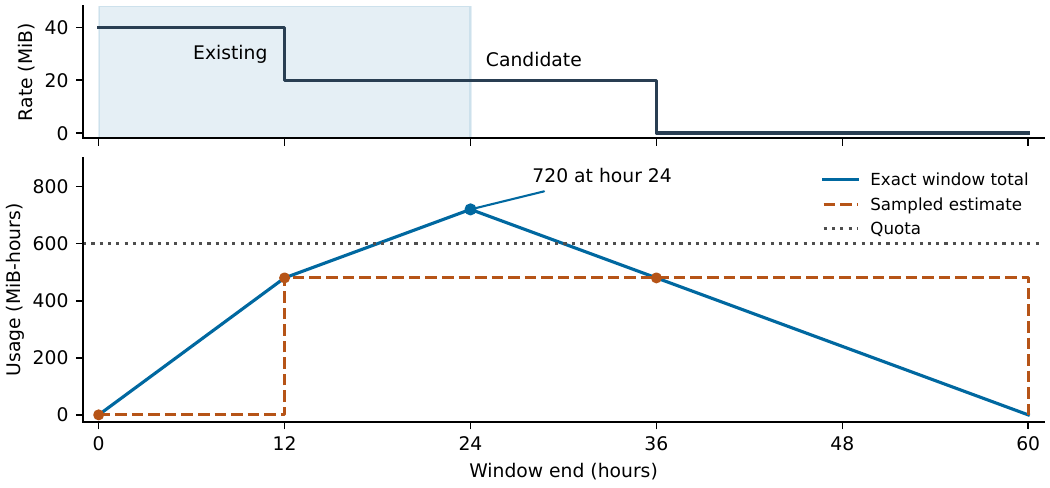}
\caption{The rate changes at hours 0, 12, and 36. The exact total peaks when the left edge reaches hour 0 and the right edge is at hour 24. At that point the exact window total is 720 MiB-hours, while the sampled estimate is 480 MiB-hours.}
\label{fig:missed-peak}
\end{figure}

A user holds 40 MiB for 12 hours and requests 20 MiB for the next 24 hours. When the window runs from hour 0 to hour 24, its left edge reaches the start of the first reservation between two sampled positions of the right edge. The window spans both rates, for a total of $12\cdot40+12\cdot20=720$ MiB-hours. The surrounding samples at hours 12 and 36 each see 480 MiB-hours. With a quota of 600 MiB-hours, the sampled policy admits the request (Figure~\ref{fig:missed-peak}).

\section{Why Sampling Fails}
\subsection{Error decomposition}
\label{sec:channels}
An earlier prefix at the left edge keeps extra usage in the window total, while an earlier prefix at the right edge leaves out recent usage. Before rounding, the difference is
\begin{equation}
H(t)-F(t)=\int_{p(t-W)}^{t-W}u(s)\,ds-\int_{p(t)}^{t}u(s)\,ds.
\label{eq:err}
\end{equation}
At a sampled right endpoint, the second term vanishes and $H(t)\ge F(t)$. A peak can still be missed if it lies between samples and exceeds every sampled estimate.

Rounding acts on the two normalized prefixes separately. In exact arithmetic with upward rounding at ties, each residual lies in $(-1/2,1/2]$. Their difference shifts the estimate by less than one average resource unit, or $W$ in resource-time units.

\subsection{Two-sided peak bounds}
\begin{prop}[Peak error on a sample-bracketed range]
\label{prop:bounds}
Consider a range whose endpoints are samples, and assume every usage change inside the range is a sample. Let $G$ be the maximum gap between consecutive samples in the range, and let $G_{\mathrm{left}}$ be the maximum of $s-W-p(s-W)$ over those samples. With $F_{\max}$ the largest exact total on the range and $H_{\max}$ the largest sampled total before rounding,
\[
F_{\max}-H_{\max}\le \frac{MG}{4},
\qquad
H_{\max}-F_{\max}\le M\,G_{\mathrm{left}}.
\]
Both constants are sharp.
\end{prop}

\begin{proof}
For consecutive samples $a<b$, write $r$ for the constant rate on $(a,b)$. Between two samples $p(t)$ is fixed while $p(t-W)$ can only advance, so $H$ is non-increasing there and its maximum over the range is attained at a sample. To bound the exact total between the samples, we can approach it from either end of the gap. The derivative satisfies $r-M\le F'(t)\le r$ almost everywhere there, so integrating forward from $a$ and backward from $b$ gives
\[
F(t)\le\min\{F(a)+r(t-a),\ F(b)+(M-r)(b-t)\}.
\]
At each sampled endpoint $F\le H$ by~\eqref{eq:err}, so $F(t)$ is bounded by $H_{\max}$ plus the smaller of $r(t-a)$ and $(M-r)(b-t)$. For $M>0$ the maximum of that smaller term is $r(M-r)(b-a)/M$, which is at most $MG/4$, and the case $M=0$ has zero error. In the other direction, at a sampled $s$ the error is at most $M(s-W-p(s-W))$ by~\eqref{eq:err}, and taking the maximum gives the second inequality.
\end{proof}

For the underestimation bound, take rates $0$, $M$ and $M/2$ on $[0,W/2)$, $[W/2,W)$ and $[W,2W)$. The exact total reaches $3MW/4$ at $3W/2$, halfway between samples. The sampled peak is $MW/2$ and the longest gap is $W$.

For the overestimation bound, use $u=M$ on $[0,W+\ell)$ with samples at $0,W,W+\ell$. At the last sample, the subtraction uses $P(0)$ and counts $M(W+\ell)$. The exact peak is $MW$ and the left lag is $\ell$. Both examples attain their respective bounds. At a clipped boundary with one neighboring sample, $|F'(t)|\le M$ bounds the additional error by $M$ times the distance to that sample.

\subsection{Non-monotonicity}
A burst can change the estimate even after it ends. When the window's left edge reaches the new sample, the policy subtracts a later prefix and drops the older usage that it had been counting. To see when this offsets the added usage, start with a rate $B<M$ on $[0,W+\ell)$ and add $\delta>0$ on $[\ell,W)$, with $B+\delta\le M$. Of the new samples $0,\ell,W,W+\ell$, the one at $\ell$ gives $B\ell$, while those at $W$ and $W+\ell$ both give $BW+\delta(W-\ell)$, which is also the exact peak. At $W+\ell$, the subtracted prefix changes from $P(0)$ to $P(\ell)$. When the added usage is smaller than the usage dropped from the left, that is, when
\[
\delta(W-\ell)<B\ell,
\]
the sampled peak falls. If the quota satisfies $BW+\delta(W-\ell)\le Q<B(W+\ell)$, the sampled policy rejects the smaller request and admits the larger one before rounding.

In the Java regression, $W=4$ seconds and $B=20$ MiB. The larger request adds $\delta=4$ MiB in the fourth second, so $\ell=3$ seconds and $M=24$ MiB. Its sample at second 3 moves the left endpoint used at second 7, bringing the sampled peak down from 140 to 84 MiB-seconds. With a quota of 100 MiB-seconds, the sampled policy rejects the smaller request and admits the larger one. The exact peaks are 80 and 84 MiB-seconds. Both requests fit.

\section{Exact Validation and Migration}
\subsection{Exact window check}
Let $v$ be the proposed rate profile: history, minus the old allocation if present, plus the candidate. For each change point $c$ of the three inputs the right rate changes at $c$ and the left rate at $c+W$, so between consecutive events
\[
F_r(t_{i+1})=F_r(t_i)+(t_{i+1}-t_i)\bigl(v_r(t_i)-v_r(t_i-W)\bigr).
\]
The scan begins by integrating the history needed for the first window. From there, it advances the integral to each event in turn, compares the total with the quota and updates the rates that change at that timestamp.

\begin{prop}[Sufficiency of endpoint checks]
\label{prop:endpoints}
On a bounded range of right endpoints, a quota violation exists if and only if the window total exceeds the quota at a range boundary or at a change point of the left or right resource rate.
\end{prop}
\begin{proof}
Between event times and range boundaries, the rates at the two ends of the window stay fixed. On each resulting interval $v(t)-v(t-W)$ is constant, so the window total is affine in $t$ and reaches its maximum at an interval endpoint. The argument applies separately in every resource dimension.
\end{proof}

The policy rejects when $F_r(t)>Q_rW$ in any dimension $r$, where $Q_r$ is the average quota in resource units after YARN's conversion. The scan reads the ordered keys once for each window endpoint. Given these maps, it takes $O(\log K+dh)$ time and $O(d)$ auxiliary space in machine-word arithmetic, where $d$ is the number of dimensions, $h$ the number of relevant points and $K$ an upper bound on the size of each map. A dimension switches to \texttt{BigInteger} if its total or limit exceeds the range of \texttt{long}.

YARN stores recurring allocations on a periodic timeline with plan period $L$. Each supported recurrence period divides $L$, so after removing the old allocation, the sum of the retained periodic rates and the recurring candidate has period $L$. For the candidate's user, each retained finite allocation on $[a,b)$ can contribute only to windows ending in $[a,b+W]$. We check the union of these ranges and a full cycle $[T,T+L]$, where $T$ is the maximum of $W$ and all finite tails $b+W$ (or $W$ if there are no finite allocations). At $T$, every finite allocation has left the window, including those with $b+W=T$.

An unchecked complete window has no finite contribution. Its total equals that of the window ending at the same phase in $[T,T+L]$, since shifting both ends by $L$ preserves the integral of the periodic rate for any $W$. At an unchecked endpoint before $W$, the window total includes only the part at or after time zero. Extending the nonnegative periodic rate backward gives a complete window that bounds this truncated total from above, and the matching phase in the checked cycle attains that bound. Range boundaries are included in the scan, which also checks windows crossing a period boundary. The implementation rejects an unsupported period or an unrepresentable range endpoint.

\subsection{Migration rule}
\label{sec:migration}
A plan admitted under the sampled policy can already exceed the exact quota. For a limit of 600 resource-hours, reducing a window from 720 to 650 removes 70 hours of excess. The migration rule admits an update if every window satisfies
\begin{equation}
(B(t)-Q)_+\le(A(t)-Q)_+\quad\Longleftrightarrow\quad B(t)\le\max\{Q,A(t)\}.
\label{eq:migration}
\end{equation}
Here $A$ and $B$ are the exact totals before and after the update. The rule preserves compliant windows and caps the excess of each violating window at its previous value. Once all windows fit, the rule gives the strict check. Every changed window lies in the union of the old and new influence ranges.

Between events, $A$ and $B$ are affine. The threshold $\max\{Q,A(t)\}$ changes slope at an interior crossing of $A(t)=Q$. Checking the segment boundaries and that crossing suffices. With endpoint totals $A_0,A_1,B_0,B_1$, the test at the crossing is the sign of
\[
|A_1-Q|(B_0-Q)+|A_0-Q|(B_1-Q).
\]
This is $B(t)-Q$ at the crossing multiplied by the positive denominator of its interpolation weights. Integer products retain the exact test. In 104{,}976 small cases checked against half-step integration, the migration rule admits 9{,}942 updates that strict validation rejects. In 1{,}553 cases, a violating window lies between integer endpoints. The trace replays use the strict policy.

\section{Feasible Starts for a Single Stage}
Consider one stage of duration $D>0$ and constant demand $d>0$ in one resource dimension. The plan is compliant and the allowed starts are the integers $s\in[a,b]$. Timestamps, $D$ and $W$ are integer ticks. For each such start, the window total is piecewise linear with integer breakpoints. Its maximum over continuous time occurs at an integer endpoint.

\subsection{Projecting violations to start intervals}
Hold a window ending at $t$ fixed and slide the stage across it. A start is forbidden when the overlap adds more usage than the slack $Q-F(t)$. The history already satisfies the quota, so $Q-F(t)\ge 0$ and the smallest violating overlap is
\[
k(t)=\left\lfloor\frac{Q-F(t)}{d}\right\rfloor+1.
\]
A window with $k(t)>\min(W,D)$ contributes no forbidden starts. Otherwise, for an overlap of at least $k(t)$, the stage has to start far enough before the right edge ($s\le t-k(t)$) and run far enough past the left edge ($s+D\ge t-W+k(t)$). Together these conditions give the closed interval of forbidden integer starts
\begin{equation}
I_t=[t-W-D+k(t),\ t-k(t)].
\label{eq:forbidden}
\end{equation}
The union of the intervals in~\eqref{eq:forbidden}, subtracted from $[a,b]$, gives the quota-feasible starts. It suffices to consider $t\in[a,b+D+W]$, since every window outside this range has zero overlap with every candidate.

\begin{prop}[Linear scan for feasible starts]
\label{prop:placement}
Given ordered changes in the usage history and in the instantaneous capacity available to the stage, all maximal feasible integer start intervals can be computed in $O(h)$ arithmetic operations and $O(h)$ space in one resource dimension. Here $h$ counts the change points read by the scan, with the two query boundaries included.
\end{prop}
\begin{proof}
Merge the changes in the historical rate and their shifts by $W$ to obtain the affine pieces of $F$. Within a piece, the integers with $F(t)>Q-d\min(W,D)$ form a contiguous range, and these are the window endpoints that give a nonempty interval in~\eqref{eq:forbidden}. To find the span of those intervals, follow their two boundaries as the window advances. Both endpoints of $I_t$ are floors or ceilings of affine functions of $t$, because $t$ is an integer and can be moved inside the floor or ceiling. Each endpoint thus moves monotonically within the piece, reaching its extreme at the first or last active integer.

To see why the intervals fill the range between those extremes, track an integer that moves one tick at a time with the window. For active $t$, the inequalities $k(t)\le\min(W,D)\le\lceil(W+D)/2\rceil\le W+D-k(t)$ place the integer $u_t=t-\lceil(W+D)/2\rceil$ inside $I_t$. Since $u_{t+1}=u_t+1$, consecutive forbidden intervals overlap or touch. Each affine piece therefore gives a single forbidden interval.

To combine the pieces, take the representative of the first active endpoint in each piece. These representatives are nondecreasing in the order the scan emits the intervals. Maintain a stack of disjoint, nonadjacent components in increasing start order, with the last representative in the top component. The right endpoint of the next interval is at least its representative, which is at least the previous representative. The interval therefore either starts beyond the top component with a gap or reaches it. In the first case, push the interval. In the second, pop the top component and extend the interval to cover both. Continue popping while the preceding component overlaps or touches the extended interval, then push the merged interval. A gap to the preceding component leaves all earlier components disjoint as well. The new top contains the new representative, preserving the invariant.

There is one push per nonempty piece, and each pushed component is popped at most once. Finding the active integer range and the two extremes takes constant arithmetic work per affine piece. The rate changes and their shifts are already ordered, so their merge produces $O(h)$ pieces in $O(h)$ time.

If instantaneous capacity is insufficient throughout a rate segment $[u,v)$, it rules out the integer starts in $[u-D+1,v-1]$. Both bounds are ordered across the rate segments, so their union and its merge with the quota components take linear time. A final scan takes the complement of the combined forbidden ranges inside $[a,b]$, yielding the maximal feasible intervals. At most $O(h)$ components are stored.
\end{proof}

For multiple dimensions, take the union of their forbidden ranges. For a scheduling grid, keep the grid points in the remaining ranges. Quota alone can split the start set. With $W=4$, $Q=8$, $D=2$ and $d=3$, a history of rate 4 on $[4,6)$ leaves starts $\{0\}\cup[8,10]$ in the query range $[0,10]$. The Python scan agrees with exhaustive enumeration in 20{,}000 random instances. Of these, 3{,}097 have several feasible intervals under the combined constraints.

\subsection{A feasible retry can displace later jobs}
A retry occupies capacity that later requests might have used, and the effect can dominate the benefit of recovering the request (Proposition~\ref{prop:retry}).
\begin{prop}[Short retries can reduce admissions]
\label{prop:retry}
For any capacity $C$ that is a positive multiple of four, with a 24-hour window and a 25\% average quota, there is a request sequence on which committing a one-second quota-feasible retry reduces the admission count from $C+1$ to 2.
\end{prop}
\begin{proof}
Let $W$ denote the window length in seconds. One user has a full-capacity reservation on $[0,W/4)$, which exhausts the 25\% quota for the first window. The user's next request asks for full capacity for one second at $W-1$, with a later start allowed. The window ending at $W$ exceeds quota at the original start, while starting at $W$ fits, because for every window spanning both allocations the old usage leaving the window is at least the new usage entering it. A retry chooses this earliest feasible start.

After this decision another user submits $C$ requests that each need one unit of capacity on $[W,W+1)$. With fixed starts all $C$ fit physically and the second user stays within quota, but after the retry the one-second slot has no capacity left. The two policies accept $C+1$ and 2 requests, including the first reservation. The retry uses $C$ core-seconds and lasts one second, so it also passes the duration and resource-time limits evaluated in Section~\ref{sec:placement-eval}.
\end{proof}

\section{Implementation}
In the Java experiments, the exact validator runs as a strict queue policy. The migration rule in Section~\ref{sec:migration} is tested separately on single-resource timelines. The feasible-start scan runs in Python for the placement experiments. To check its output against the ResourceManager, we submit the computed boundary starts as fixed-start requests under the strict policy.

The Java harness uses Hadoop 3.5.0 binaries, built from release commit \href{https://github.com/apache/hadoop/commit/dbcc7cd797100e6b32cd84f85b53a5193a5f9af0}{\texttt{dbcc7cd}}. For the sampled baseline, we compile \texttt{CapacityOverTimePolicy} from the YARN-12022 proposal, PR 8815 at \href{https://github.com/apache/hadoop/commit/f99af959fd4e26cf0b359e6f341d560761d14101}{\texttt{f99af95}}~\cite{yarn12022}. The unpatched reproducer uses the policy from trunk commit \href{https://github.com/apache/hadoop/commit/1fdf777d777d74e9de04961e9a657e61107eadec}{\texttt{1fdf777}} with the same harness. YARN-12022 bounds the merge horizon to avoid overflow and removes the old allocation from the accounting for a replacement. Its integral check still subtracts rounded sampled prefixes.

The exact queue policy retains YARN's instantaneous check and integrates the plan's RLE timeline at changes at either window edge. A replacement can reach validation twice, once before YARN removes the old allocation and again on write-back. The validator looks up the reservation ID on each call and subtracts the old allocation if the plan still contains it.

Both pinned versions of the sampled policy admit the over-quota request in Section~\ref{sec:missed-peak}, which the exact policy rejects. In two boundary regressions, sustaining the quota rate for 25 hours fits every 24-hour window. The exact policy admits both cases.

\section{Evaluation}
\label{sec:eval}
\subsection{Workload and replay design}
We use the cleaned KTH SP2 log from the Parallel Workloads Archive~\cite{kth,feitelsonarchive} and the grid-job records from the CERIT-SC study~\cite{cerit,ceritstudy}. We retain completed jobs with nonnegative submission and wait times, positive runtime, a user identifier and a processor count within plan capacity. We order eligible jobs by submission time and identifier. Each replay takes the first 1{,}450 records, with 250 for warm-up and 1{,}200 for measurement. The KTH slice contains 64 users and runs on a 100-vCore plan. The CERIT slice contains 37 users and runs on a 256-vCore plan.

For each accounting record, the reservation starts at the submission time plus the recorded wait. Its duration and vCore count come from the recorded execution, with 1{,}024 MiB assigned per vCore. We test configured average quotas of 2\%, 5\%, 10\% and 25\% over 24 hours and use a 100\% instantaneous limit. We measure excess and headroom against the effective quota after YARN's resource-unit conversion. Both policies see identical histories in same-state replay, while in continuous replay each retains its own accepted allocations.

\subsection{Admission disagreements}
For each request, we call the two policies in random order. If they disagree, we keep the exact decision in both plans so the next request sees the same history. Including warm-up, independent integration agrees with the exact policy on all 10{,}075 requests that reach the integral check. The remaining 1{,}525 fail the instantaneous check under both policies. Table~\ref{tab:same} reports the 1{,}200 measured requests per configuration.

In KTH, all 93 disagreements at a 2\% quota are false admissions, while all five at a 5\% quota are false rejections. CERIT 10\% has the highest disagreement rate, with 123 false admissions and 36 false rejections in 1{,}200 requests (Table~\ref{tab:same}). In the CERIT 5\% replay, the 95th percentile of excess among the 54 false admissions is 44.2\% of quota. Among falsely rejected requests in CERIT 10\%, the largest headroom is 83.1\% of quota.

\begin{table}[t]
\centering
\caption{Same-state decisions for 1{,}200 measured requests per row. FA denotes a false admission by the sampled policy and FR a false rejection. Both policies share the instantaneous denials. The two rate columns use all 1{,}200 requests.}
\label{tab:same}
\small
\begin{tabular}{lrrrrrr}
\toprule
Trace & Quota & Instantaneous denials & Exact integral rejection & FA & FR & Disagreement \\
\midrule
KTH    & 2\%  & 0   & 20.1\% & 93 & 0  & 7.8\% \\
KTH    & 5\%  & 0   & 6.3\%  & 0  & 5  & 0.4\% \\
KTH    & 10\% & 0   & 2.1\%  & 0  & 0  & 0\% \\
KTH    & 25\% & 0   & 0.5\%  & 0  & 0  & 0\% \\
CERIT  & 2\%  & 0   & 78.2\% & 68 & 8  & 6.3\% \\
CERIT  & 5\%  & 2   & 70.2\% & 54 & 29 & 6.9\% \\
CERIT  & 10\% & 365 & 47.0\% & 123& 36 & 13.3\% \\
CERIT  & 25\% & 919 & 3.6\%  & 0  & 13 & 1.1\% \\
\bottomrule
\end{tabular}
\end{table}

\subsection{Rounding versus sample spacing}
\label{sec:decomposition}
We give four Python validators the same history from the exact plan so we can vary rounding and sample spacing separately. We compare rounded and exact prefixes at spacing $W$, then keep rounding and use spacings $W/2$ and $W/4$. Each configuration starts with 1{,}450 requests, including warm-up, and Table~\ref{tab:decomposition} counts those that reach the integral check. In continuous replay, the Python reference reproduces the Java exact verdicts and the Java sampled admission totals for all eight configurations.

\begin{table}[t]
\centering
\small
\caption{False admissions (FA) and false rejections (FR) with the exact policy's history. Sampled and unrounded both use spacing $W$. The finer grids retain prefix rounding. Checks count requests that pass the instantaneous check.}
\label{tab:decomposition}
\begin{tabular}{@{}lrrrrrrrrrr@{}}
\toprule
& & & \multicolumn{2}{c}{Sampled} & \multicolumn{2}{c}{Unrounded} & \multicolumn{2}{c}{$W/2$} & \multicolumn{2}{c}{$W/4$} \\
\cmidrule(lr){4-5}\cmidrule(lr){6-7}\cmidrule(lr){8-9}\cmidrule(l){10-11}
Trace, quota & Checks & Exact admits & FA & FR & FA & FR & FA & FR & FA & FR \\
\midrule
KTH, 2\%    & 1{,}450 & 1{,}163 & 107 & 0 & 0 & 2 & 107 & 0 & 107 & 0 \\
KTH, 5\%    & 1{,}450 & 1{,}363 & 0 & 7 & 0 & 7 & 0 & 6 & 0 & 3 \\
KTH, 10\%   & 1{,}450 & 1{,}423 & 0 & 0 & 0 & 0 & 0 & 0 & 0 & 0 \\
KTH, 25\%   & 1{,}450 & 1{,}444 & 0 & 0 & 0 & 0 & 0 & 0 & 0 & 0 \\
CERIT, 2\%  & 1{,}450 & 283 & 69 & 9 & 9 & 11 & 68 & 7 & 66 & 2 \\
CERIT, 5\%  & 1{,}448 & 392 & 93 & 33 & 66 & 37 & 81 & 24 & 73 & 22 \\
CERIT, 10\% & 1{,}032 & 316 & 129 & 46 & 96 & 53 & 17 & 20 & 0 & 20 \\
CERIT, 25\% & 345 & 280 & 0 & 18 & 0 & 18 & 0 & 14 & 0 & 12 \\
\bottomrule
\end{tabular}
\end{table}

The effective CPU quota in KTH 2\% is one vCore, on the scale of the rounding error. Removing rounding eliminates all 107 false admissions. Both denser grids retain all 107 with rounding enabled. In CERIT 10\%, exact arithmetic at spacing $W$ leaves 96 false admissions and 53 false rejections, while the $W/4$ grid with rounded prefixes eliminates false admissions and leaves 20 false rejections. CERIT 25\% still has 18 false rejections after removing rounding. The $W/4$ grid brings that count down to 12.

\subsection{Continuous replay}
\begin{table}[t]
\centering
\caption{Continuous replay of 1{,}450 requests per configuration. The violation rate is the share of sampled-policy admissions with an over-quota window in the candidate's influence range. The last two columns show the percentage by which the largest window total in each accepted history differs from quota. Negative values denote headroom.}
\label{tab:continuous}
\small
\begin{tabular}{lrrrrrr}
\toprule
Trace, quota & Sampled admits & Exact admits & Violating & Rate & Sampled history & Exact history \\
\midrule
KTH, 2\%    & 1{,}194 & 1{,}163 & 79 & 6.6\% & 71.8\% & 0\% \\
KTH, 5\%    & 1{,}357 & 1{,}363 & 0  & 0\%   & $-1.9$\% & $-0.4$\% \\
KTH, 10\%   & 1{,}423 & 1{,}423 & 0  & 0\%   & $-1.0$\% & $-1.0$\% \\
KTH, 25\%   & 1{,}444 & 1{,}444 & 0  & 0\%   & $-3.3$\% & $-3.3$\% \\
CERIT, 2\%  & 279 & 283 & 18 & 6.5\% & 32.1\% & $-0.1$\% \\
CERIT, 5\%  & 380 & 392 & 11 & 2.9\% & 44.6\% & 0\% \\
CERIT, 10\% & 328 & 316 & 4  & 1.2\% & 12.0\% & $-0.2$\% \\
CERIT, 25\% & 327 & 280 & 0  & 0\%   & $-6.6$\% & 0\% \\
\bottomrule
\end{tabular}
\end{table}

In KTH 2\%, 79 of 1{,}194 sampled admissions overlap an over-quota window, and 16 of the 64 users have at least one over-quota window. The peak is 71.8\% above the effective quota of 1 vCore averaged over 24 hours. In CERIT 5\%, the sampled history peaks at 44.6\% above quota. The exact policy stays within quota in all eight configurations.

The sampled and exact policies admit 328 and 316 jobs in CERIT 10\%. Those totals are close even though the policies differ on 184 of the 1{,}450 individual decisions.

\subsection{Quota conversion}
YARN converts percentages to resource units with \texttt{Resources.multiply(Resource, float)} and truncates each limit. The stored value of \texttt{0.02f} lies just below 0.02, so a 2\% quota on the 100-vCore plan gives 1 vCore and 2{,}047 MiB. The decimal products are 2 vCores and 2{,}048 MiB. Both policies use the same YARN limits in Table~\ref{tab:continuous}.

We reran the continuous replay with the Python reference policies, starting each history empty and replacing each average limit by $\lfloor pC/100+1/2\rfloor$ for a configured percentage $p$ and resource capacity $C$. The replay first reproduces the admission counts and peak excesses in Table~\ref{tab:continuous} with the original limits. After the change, the peak excess in the sampled history drops from 71.8\% to 0.2\% in KTH 2\%. It remains 40.7\% in CERIT 5\%, down from 44.6\%, and 9.4\% in CERIT 10\%, down from 12.0\%. The exact histories stay within the rounded limits in all eight configurations. Exact admissions rise from 1{,}163 to 1{,}228, from 392 to 398, and from 316 to 324 in the three cases, respectively.

\subsection{Multiple resources, replacements and recurrence}
By varying memory from 512 to 4{,}096 MiB per vCore, we let either resource reach its limit independently of the other. In the CERIT 5\% replay, memory binds in 37\% of quota checks, and the sampled policy produces 57 false admissions and 25 false rejections on the exact policy's history.

We replace every tenth admitted reservation with one that lasts twice as long and uses two more vCores. The sampled policy falsely admits 10 of the 117 replacements checked in KTH 2\%. Recurring allocations can contribute usage from both sides of a period boundary, and the validator agrees with direct integration on all 64 cases spanning a boundary of the 24-hour period.

\subsection{Plan-entry and validation latency}
We collected the Java timings on Windows with an Intel Core i9-14900HX and JDK 26.0.1. The timer for \texttt{addReservation} covers plan reads, validation and write-back. Request construction, archiving and recording run outside the timer. For requests that both policies admit, we take each request's median across three JVM runs and report the 95th percentile of those medians.

\begin{table}[t]
\centering
\caption{Plan-entry P95 latency in $\mu$s for requests admitted by both policies. Each request contributes its median across three JVM runs.}
\label{tab:latency}
\begin{tabular}{llrr}
\toprule
Trace, quota & Paired admissions & Sampled & Exact \\
\midrule
KTH, 2\%    & 959    & 446.3 & 185.5 \\
KTH, 5\%    & 1{,}119 & 397.5 & 154.4 \\
KTH, 10\%   & 1{,}175 & 449.4 & 175.3 \\
KTH, 25\%   & 1{,}194 & 217.3 & 87.4 \\
CERIT, 2\%  & 253    & 589.5 & 240.8 \\
CERIT, 5\%  & 327    & 924.8 & 371.8 \\
CERIT, 10\% & 235    & 236.8 & 96.3 \\
CERIT, 25\% & 225    & 182.7 & 85.3 \\
\bottomrule
\end{tabular}
\end{table}

Across the eight configurations, exact plan-entry P95 is 39--47\% of sampled plan-entry P95. In the CERIT 5\% replay, this corresponds to a drop from 924.8 to 371.8 $\mu$s (Table~\ref{tab:latency}).

To vary the number of events independently of request duration, we fill a 4{,}096-second history with successively shorter rate segments and keep the candidate fixed. Both policies admit it under a 24-hour window and a 25\% quota.

\begin{figure}[tb]
\centering
\includegraphics[width=\linewidth]{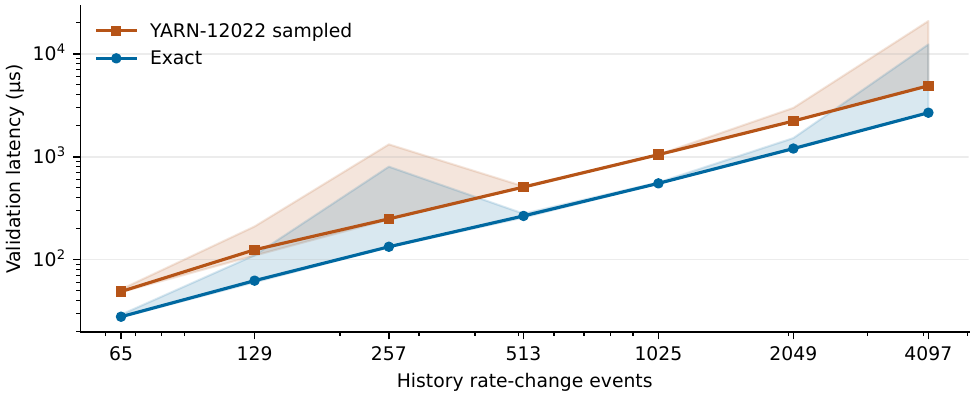}
\caption{Policy validation latency on accepted requests with fixed duration and varying history event counts. Shading spans the minimum and maximum of the three per-JVM medians.}
\label{fig:event-scaling}
\end{figure}

We time only the paired \texttt{validate} calls, randomize their order, and collect 200 pairs per event count after 150 warm-up pairs in each of three JVMs. Figure~\ref{fig:event-scaling} plots the median of the three per-JVM medians. The exact validator has a lower median at every tested size. At 4{,}097 history events, the medians are 2.67 ms for the exact validator and 4.89 ms for sampling.

\subsection{End-to-end RPC latency}
We reproduce the missed peak through the ResourceManager's submission and update RPCs. In a 100-vCore plan, the first 24-hour window reaches 720 vCore-hours against a limit of 600. The sampled policy admits the request. Exact validation rejects it and, for the update, leaves the previous 5-vCore allocation in place.

We configure two equal-capacity queues in a ResourceManager with a simulated node and in-memory plan state. Each pair of calls starts from the same history. Each of four JVM runs measures 400 pairs of successful submissions and 400 pairs of successful updates after 150 warm-up pairs per operation. We randomize policy order within each pair and time the full client call over loopback. The median of the four P95s is 0.433 ms for sampled and 0.426 ms for exact submission. For update, the corresponding medians are 0.445 and 0.437 ms. In one run the sampled policy has lower P95 for both calls.

\section{Flexible Placement}
\label{sec:placement-eval}
\subsection{A feasible placement missed by the default agent}
In the 100-vCore plan, a user holds 40 vCores during hours $[0,12)$ and $[52,64)$. The user asks for 20 vCores for 24 hours, arriving at hour 12 with a deadline at hour 52. Under exact validation, the default agent's choice exceeds the integral quota, while the scan finds feasible starts at hours 18 through 22. ResourceManager accepts fixed-start RPCs at both boundaries and rejects them one scheduling step outside.

\subsection{Recoverable requests}
After trying the recorded start, we give integral rejections a synthetic deadline and test later starts against the same unchanged history. Within four hours, 233 requests across eight configurations have another feasible start. At CERIT 5\%, this covers 136 of the 842 integral rejections. On 195 requests that fit in isolation, the interval scan and cached enumeration agree on a 60-second grid, with Python search P95s of 160 and 244 $\mu$s, respectively, for a one-hour search range.

\subsection{Committed retries}
The fixed-start Python baseline reproduces all 11{,}600 exact Java decisions, including warm-up. In the retry run, a request that passes the instantaneous check and exceeds the average quota can move to the earliest feasible alternative, which becomes part of the history for later requests. With a four-hour budget, admissions fall from 356 to 345 at CERIT 5\% and from 5{,}578 to 5{,}564 across the original eight configurations.

\subsection{Bounded retry policies}
We take the next three disjoint slices from each source, start each from an empty plan, use 250 warm-up jobs and measure the next 1{,}200 at each quota. We compare unrestricted one- and four-hour retries against three four-hour rules that cap the stage at 15 minutes, at one hour, or at one hour plus a resource-time cap equal to the full plan for 15 minutes. A request is short if it lasts at most one hour.

On the later slices, unrestricted four-hour retry adds 261 net admissions. Twelve configurations gain admissions and three lose them. Limiting retries to stages of at most one hour yields 215 net admissions and 229 net short admissions. Compared with unrestricted retry, the duration cap admits 95 more short requests and 46 fewer requests overall. The median delay among shifted requests falls from 185.7 to 43.4 minutes. On the original slices, the duration cap loses two admissions relative to fixed starts (Table~\ref{tab:retry-guards}).

\begin{table}[t]
\centering
\caption{Net admissions relative to fixed starts and delay among shifted requests. The fixed policy admits 5{,}578 of 9{,}600 measured requests on the original slices and 15{,}913 of 28{,}800 on the later slices. The later slices cover 24 configurations.}
\label{tab:retry-guards}
\begin{tabular}{@{}lrrrrr@{}}
\toprule
& Original & \multicolumn{2}{c}{Later net admissions} & \multicolumn{2}{c}{Later delay (min)} \\
\cmidrule(lr){3-4}\cmidrule(l){5-6}
Retry rule & Net admissions & All & Short & Median & P95 \\
\midrule
Unrestricted, 1 h budget & $-2$ & $+151$ & $+80$ & 42.9 & 56.1 \\
Unrestricted, 4 h budget & $-14$ & $+261$ & $+134$ & 185.7 & 225.1 \\
Stage $\le$15 min & $+3$ & $+19$ & $+18$ & 61.8 & 156.0 \\
Stage $\le$1 h & $-2$ & $+215$ & $+229$ & 43.4 & 213.7 \\
Stage $\le$1 h with resource-time cap & $-2$ & $+215$ & $+229$ & 43.4 & 213.7 \\
\bottomrule
\end{tabular}
\end{table}

\section{Related Work}
Rayon~\cite{rayon} introduced YARN's reservation language and separated placement from validation against resource and sharing constraints. Its technical report describes checking the quota of a fixed allocation in linear time by sliding a window over the user's reservations and updating the integral at each change~\cite[Sec.~4.1]{rayonreport}. YARN-1711 added the average-capacity policy~\cite{yarn1711}, which checks the concrete allocation returned by the reservation agent~\cite{planning}. YARN-12022 proposes bounding the integration range to avoid overflow and subtracting the old allocation on update~\cite{yarn12022}. It retains sampled prefixes. We analyze the errors of that computation, test the exact window scan in Hadoop 3.5.0, and use window constraints to compute all feasible starts of a single stage.

Datar et al.~\cite{datarwindows} give error guarantees for summaries of the most recent stream items. The Reactive Aggregator updates window aggregates as items change, and FiBA supports out-of-order insertions and evictions~\cite{tangwongsanra,tangwongsanfiba}. At admission, YARN already has a timeline of committed rates. It must decide whether adding a proposed allocation would put any affected window over quota.

Chen and Lee~\cite{chenflexible} give an advance reservation a range of possible starts, and Li et al.~\cite{liavailability} index processor availability in slots to place jobs with deadlines. The intervals we compute account for the user's integral quota as well as physical availability. TetriSched updates a global plan for pending jobs~\cite{tetrisched}, while Mu'alem and Feitelson~\cite{mualem} compare utilization and predictability under backfilling on the IBM SP2. In our replay, existing placements stay fixed as we trace how admitting a request at an alternative start changes later decisions.

\section{Conclusion}
The sampled YARN policy can admit an allocation that exceeds quota and reject one that fits. The exact policy stays within quota in every continuous replay, and its plan-entry latency is lower on the requests both policies admit. A scan of the same timeline finds where a single stage can run. Taking one of those starts can displace later jobs. On the later trace slices, capping stage duration at one hour admits more short jobs and reduces median delay among shifted requests compared with unrestricted four-hour retries.

\section*{Acknowledgements}
The author thanks the maintainers of the public workload archives for making the KTH SP2 and CERIT-SC traces available.

\section*{Use of Generative AI}
OpenAI Codex assisted with editing the manuscript and scripting the quota-conversion sensitivity replay. The replay uses the public trace inputs and reference validators in the artifact below.

\section*{Conflict of Interest}
The author declares no conflict of interest.

\section*{Data Availability Statement}
The policy sources, replay scripts, derived reservation streams and measurement summaries are available in the \href{https://github.com/stormwhip/SlidingWindowQuota/tree/534c415191a02450c767c31e0d95889aa1320e55}{SlidingWindowQuota artifact (commit \texttt{534c415})}. The download script verifies the SHA-256 digests of the two public accounting traces.

\bibliographystyle{plainnat}
\bibliography{refs}

@inproceedings{vavilapalli,
  author = {Vinod Kumar Vavilapalli and Arun C. Murthy and Chris Douglas and
            Sharad Agarwal and Mahadev Konar and Robert Evans and Thomas Graves
            and Jason Lowe and Hitesh Shah and Siddharth Seth and Bikas Saha and
            Carlo Curino and Owen O'Malley and Sanjay Radia and Benjamin Reed
            and Eric Baldeschwieler},
  title = {{Apache Hadoop YARN}: Yet Another Resource Negotiator},
  booktitle = {Proceedings of the ACM Symposium on Cloud Computing},
  year = {2013},
  pages = {5:1--5:16},
  doi = {10.1145/2523616.2523633}
}

@inproceedings{rayon,
  author = {Carlo Curino and Djellel E. Difallah and Chris Douglas and
            Subru Krishnan and Raghu Ramakrishnan and Sriram Rao},
  title = {Reservation-Based Scheduling: If You're Late Don't Blame Us!},
  booktitle = {Proceedings of the ACM Symposium on Cloud Computing},
  year = {2014},
  pages = {1--14},
  doi = {10.1145/2670979.2670981}
}

@techreport{rayonreport,
  author = {Carlo Curino and Djellel E. Difallah and Chris Douglas and
            Subru Krishnan and Raghu Ramakrishnan and Sriram Rao},
  title = {Reservation-Based Scheduling: If You're Late Don't Blame Us!},
  institution = {Microsoft Research},
  number = {MSR-TR-2013-108},
  year = {2013},
  url = {https://www.microsoft.com/en-us/research/wp-content/uploads/2013/10/rayon.pdf}
}

@misc{yarn1711,
  author = {{Apache Hadoop}},
  title = {{YARN-1711}: {CapacityOverTimePolicy}, a Policy to Enforce Quotas over Time},
  year = {2014},
  howpublished = {\href{https://issues.apache.org/jira/browse/YARN-1711}{Issue tracker}}
}

@misc{yarn12022,
  author = {{Apache Hadoop}},
  title = {{YARN-12022}: Fix {CapacityOverTimePolicy} Rejecting Reservations and Updates within the Quota},
  year = {2026},
  howpublished = {\href{https://github.com/apache/hadoop/pull/8815}{Pull request 8815}}
}

@misc{capacitydocs,
  author = {{Apache Hadoop}},
  title = {{CapacityScheduler} Documentation, Release 3.5.0},
  year = {2026},
  howpublished = {\href{https://hadoop.apache.org/docs/r3.5.0/hadoop-yarn/hadoop-yarn-site/CapacityScheduler.html}{Documentation}}
}

@misc{hadoopres,
  author = {{Apache Hadoop}},
  title = {Reservation System Documentation, Release 3.5.0},
  year = {2026},
  howpublished = {\href{https://hadoop.apache.org/docs/r3.5.0/hadoop-yarn/hadoop-yarn-site/ReservationSystem.html}{Documentation}}
}

@misc{planning,
  author = {{Apache Hadoop}},
  title = {{PlanningAlgorithm}, Release 3.5.0},
  year = {2026},
  howpublished = {\href{https://github.com/apache/hadoop/blob/rel/release-3.5.0/hadoop-yarn-project/hadoop-yarn/hadoop-yarn-server/hadoop-yarn-server-resourcemanager/src/main/java/org/apache/hadoop/yarn/server/resourcemanager/reservation/planning/PlanningAlgorithm.java}{Source code}}
}

@misc{kth,
  author = {Lars Malinowsky},
  title = {{KTH SP2} Workload Trace},
  year = {1996--1997},
  howpublished = {\href{https://www.cs.huji.ac.il/labs/parallel/workload/l_kth_sp2/}{Parallel Workloads Archive}}
}

@misc{cerit,
  author = {Dalibor Klus{\'a}{\v c}ek and Boris Par{\'a}k},
  title = {{CERIT-SC} Workload Traces},
  year = {2017},
  howpublished = {\href{https://jsspp.org/workload/index.php?page=cerit}{JSSPP Workload Archive}}
}

@article{feitelsonarchive,
  author = {Dror G. Feitelson and Dan Tsafrir and David Krakov},
  title = {Experience with Using the Parallel Workloads Archive},
  journal = {Journal of Parallel and Distributed Computing},
  year = {2014},
  volume = {74},
  number = {10},
  pages = {2967--2982},
  doi = {10.1016/j.jpdc.2014.06.013}
}

@inproceedings{ceritstudy,
  author = {Dalibor Klus{\'a}{\v c}ek and Boris Par{\'a}k},
  title = {Analysis of Mixed Workloads from Shared Cloud Infrastructure},
  booktitle = {Job Scheduling Strategies for Parallel Processing},
  series = {Lecture Notes in Computer Science},
  volume = {10773},
  year = {2018},
  pages = {25--42},
  doi = {10.1007/978-3-319-77398-8_2}
}

@article{datarwindows,
  author = {Mayur Datar and Aristides Gionis and Piotr Indyk and Rajeev Motwani},
  title = {Maintaining Stream Statistics over Sliding Windows},
  journal = {SIAM Journal on Computing},
  year = {2002},
  volume = {31},
  number = {6},
  pages = {1794--1813},
  doi = {10.1137/S0097539701398363}
}

@article{tangwongsanra,
  author = {Kanat Tangwongsan and Martin Hirzel and Scott Schneider and Kun-Lung Wu},
  title = {General Incremental Sliding-Window Aggregation},
  journal = {Proceedings of the VLDB Endowment},
  year = {2015},
  volume = {8},
  number = {7},
  pages = {702--713},
  doi = {10.14778/2752939.2752940}
}

@article{tangwongsanfiba,
  author = {Kanat Tangwongsan and Martin Hirzel and Scott Schneider},
  title = {Optimal and General Out-of-Order Sliding-Window Aggregation},
  journal = {Proceedings of the VLDB Endowment},
  year = {2019},
  volume = {12},
  number = {10},
  pages = {1167--1180},
  doi = {10.14778/3339490.3339499}
}

@article{chenflexible,
  author = {Y. T. Chen and K. H. Lee},
  title = {A Flexible Service Model for Advance Reservation},
  journal = {Computer Networks},
  year = {2001},
  volume = {37},
  number = {3--4},
  pages = {251--262},
  doi = {10.1016/S1389-1286(01)00191-8}
}

@article{liavailability,
  author = {Bo Li and Yijian Pei and Hao Wu and Bin Shen},
  title = {Resource Availability-Aware Advance Reservation for Parallel Jobs with Deadlines},
  journal = {The Journal of Supercomputing},
  year = {2014},
  volume = {68},
  number = {2},
  pages = {798--819},
  doi = {10.1007/s11227-013-1067-8}
}

@inproceedings{tetrisched,
  author = {Alexey Tumanov and Timothy Zhu and Jun Woo Park and Michael A. Kozuch
            and Mor Harchol-Balter and Gregory R. Ganger},
  title = {{TetriSched}: Global Rescheduling with Adaptive Plan-Ahead in Dynamic Heterogeneous Clusters},
  booktitle = {Proceedings of the European Conference on Computer Systems},
  year = {2016},
  pages = {35:1--35:16},
  doi = {10.1145/2901318.2901355}
}

@article{mualem,
  author = {Ahuva W. Mu'alem and Dror G. Feitelson},
  title = {Utilization, Predictability, Workloads, and User Runtime Estimates in Scheduling the {IBM SP2} with Backfilling},
  journal = {IEEE Transactions on Parallel and Distributed Systems},
  year = {2001},
  volume = {12},
  number = {6},
  pages = {529--543},
  doi = {10.1109/71.932708}
}

\end{document}